\documentstyle[10pt,epsf,epsfig,dp_delphititle]{dp_delphi}
\makeindex
\def\DpPaperGroup{EP}
\def\DpPaperRef{98--177}
\def\DpDate{6 November 1998}
\def\DpAuthors{DELPHI Collaboration}
\def\DpSubmit{(Submitted to Physics Letters B)}
\def\DpTitle{{Search for Leptoquarks and FCNC in \boldmath $e^+e^-$ annihilations at $\sqrt{s}=$183~GeV}}

\begin{document}
%%%%%%%%%%%%%%%%%%%%%%%%%% They are a problem with Coll.Sty ?
\makeatletter
% Collapse citation numbers to ranges.  Non-numeric and undefined labels
% are handled.  No sorting is done.  E.g., 1,3,2,3,4,5,foo,1,2,3,?,4,5
% gives 1,3,2-5,foo,1-3,?,4,5
\newcount\@tempcntc
\def\@citex[#1]#2{\if@filesw\immediate\write\@auxout{\string\citation{#2}}\fi
  \@tempcnta\z@\@tempcntb\m@ne\def\@citea{}\@cite{\@for\@citeb:=#2\do
    {\@ifundefined
       {b@\@citeb}{\@citeo\@tempcntb\m@ne\@citea\def\@citea{,}{\bf ?}\@warning
       {Citation `\@citeb' on page \thepage \space undefined}}%
    {\setbox\z@\hbox{\global\@tempcntc0\csname b@\@citeb\endcsname\relax}%
     \ifnum\@tempcntc=\z@ \@citeo\@tempcntb\m@ne
       \@citea\def\@citea{,}\hbox{\csname b@\@citeb\endcsname}%
     \else
      \advance\@tempcntb\@ne
      \ifnum\@tempcntb=\@tempcntc
      \else\advance\@tempcntb\m@ne\@citeo
      \@tempcnta\@tempcntc\@tempcntb\@tempcntc\fi\fi}}\@citeo}{#1}}
\def\@citeo{\ifnum\@tempcnta>\@tempcntb\else\@citea\def\@citea{,}%
  \ifnum\@tempcnta=\@tempcntb\the\@tempcnta\else
   {\advance\@tempcnta\@ne\ifnum\@tempcnta=\@tempcntb \else \def\@citea{--}\fi
    \advance\@tempcnta\m@ne\the\@tempcnta\@citea\the\@tempcntb}\fi\fi}
 
\makeatother
%%%%%%%%%%%%%%%%%%%%%%%%%% ??????????????????????????????????
% Generate the title page
\begin{titlepage}
\pagenumbering{roman}
\CERNpreprint{\DpPaperGroup}{\DpPaperRef} % Reference of the paper
\date{{\small\DpDate}} % Date of the paper
\title{\DpTitle} % Title of the paper
\address{\DpAuthors} % General name of the author(s)
\begin{shortabs} % Start the abstract
\noindent
%   abstract.tex
%
\noindent

A search for events with one jet and at most one isolated lepton
used data taken at LEP-2 by the DELPHI detector.
These data were accumulated at a center-of-mass energy of
183 GeV and correspond to an integrated luminosity of 
47.7 pb$^{-1}$.
Production of single scalar and vector leptoquarks 
was searched for. Limits at 95\% confidence level were 
derived on the masses (ranging from $134~{\rm GeV/c^2}$ to $171~{\rm GeV/c^2}$
 for electromagnetic type couplings) and couplings of the leptoquark states.
A search for top-charm flavour changing neutral currents 
($e^+ e^- \rightarrow \bar t c$ or charge conjugate)
 used the semileptonic decay channel. A limit on the flavour changing cross-section via
neutral currents was set at 0.55pb (95\% confidence level).
\end{shortabs}
\vfill
\begin{center}
\DpSubmit \ % Horrible hack to allow to have DpSubmit empty
%\DpComment \ \\
%\DpEMail \ \\
\end{center}
\vfill
%\clearpage
\headsep 10.0pt
\begingroup
% Commands to process the author names
%
\newcommand{\DpName}[2]{\hbox{#1$^{\ref{#2}}$},\hfill}
\newcommand{\DpNameTwo}[3]{\hbox{#1$^{\ref{#2},\ref{#3}}$},\hfill}
\newcommand{\DpNameThree}[4]{\hbox{#1$^{\ref{#2},\ref{#3},\ref{#4}}$},\hfill}
\newskip\Bigfill \Bigfill = 0pt plus 1000fill
\newcommand{\DpNameLast}[2]{\hbox{#1$^{\ref{#2}}$}\hspace{\Bigfill}}
\small
%%%CD\footnotesize
\pagenumbering{roman}
\addtocounter{page}{-1}
\noindent
\DpName{P.Abreu}{LIP}
\DpName{W.Adam}{VIENNA}
\DpName{T.Adye}{RAL}
\DpName{P.Adzic}{DEMOKRITOS}
\DpName{T.Aldeweireld}{AIM}
\DpName{G.D.Alekseev}{JINR}
\DpName{R.Alemany}{VALENCIA}
\DpName{T.Allmendinger}{KARLSRUHE}
\DpName{P.P.Allport}{LIVERPOOL}
\DpName{S.Almehed}{LUND}
% ...
\titlefoot{Department of Physics and Astronomy, Iowa State
     University, Ames IA 50011-3160, USA
    \label{AMES}}
\titlefoot{Physics Department, Univ. Instelling Antwerpen,
     Universiteitsplein 1, BE-2610 Wilrijk, Belgium \\
     \indent~~and IIHE, ULB-VUB,
     Pleinlaan 2, BE-1050 Brussels, Belgium \\
     \indent~~and Facult\'e des Sciences,
     Univ. de l'Etat Mons, Av. Maistriau 19, BE-7000 Mons, Belgium
    \label{AIM}}
\titlefoot{Physics Laboratory, University of Athens, Solonos Str.
     104, GR-10680 Athens, Greece
    \label{ATHENS}}
\titlefoot{Department of Physics, University of Bergen,
     All\'egaten 55, NO-5007 Bergen, Norway
    \label{BERGEN}}
\titlefoot{Dipartimento di Fisica, Universit\`a di Bologna and INFN,
     Via Irnerio 46, IT-40126 Bologna, Italy
    \label{BOLOGNA}}
\titlefoot{Centro Brasileiro de Pesquisas F\'{\i}sicas, rua Xavier Sigaud 150,
     BR-22290 Rio de Janeiro, Brazil \\
     \indent~~and Depto. de F\'{\i}sica, Pont. Univ. Cat\'olica,
     C.P. 38071 BR-22453 Rio de Janeiro, Brazil \\
     \indent~~and Inst. de F\'{\i}sica, Univ. Estadual do Rio de Janeiro,
     rua S\~{a}o Francisco Xavier 524, Rio de Janeiro, Brazil
    \label{BRASIL}}
\titlefoot{Comenius University, Faculty of Mathematics and Physics,
     Mlynska Dolina, SK-84215 Bratislava, Slovakia
    \label{BRATISLAVA}}
\titlefoot{Coll\`ege de France, Lab. de Physique Corpusculaire, IN2P3-CNRS,
     FR-75231 Paris Cedex 05, France
    \label{CDF}}
\titlefoot{CERN, CH-1211 Geneva 23, Switzerland
    \label{CERN}}
\titlefoot{Institut de Recherches Subatomiques, IN2P3 - CNRS/ULP - BP20,
     FR-67037 Strasbourg Cedex, France
    \label{CRN}}
\titlefoot{Institute of Nuclear Physics, N.C.S.R. Demokritos,
     P.O. Box 60228, GR-15310 Athens, Greece
    \label{DEMOKRITOS}}
\titlefoot{FZU, Inst. of Phys. of the C.A.S. High Energy Physics Division,
     Na Slovance 2, CZ-180 40, Praha 8, Czech Republic
    \label{FZU}}
\titlefoot{Dipartimento di Fisica, Universit\`a di Genova and INFN,
     Via Dodecaneso 33, IT-16146 Genova, Italy
    \label{GENOVA}}
\titlefoot{Institut des Sciences Nucl\'eaires, IN2P3-CNRS, Universit\'e
     de Grenoble 1, FR-38026 Grenoble Cedex, France
    \label{GRENOBLE}}
\titlefoot{Helsinki Institute of Physics, HIP,
     P.O. Box 9, FI-00014 Helsinki, Finland
    \label{HELSINKI}}
\titlefoot{Joint Institute for Nuclear Research, Dubna, Head Post
     Office, P.O. Box 79, RU-101 000 Moscow, Russian Federation
    \label{JINR}}
\titlefoot{Institut f\"ur Experimentelle Kernphysik,
     Universit\"at Karlsruhe, Postfach 6980, DE-76128 Karlsruhe,
     Germany
    \label{KARLSRUHE}}
\titlefoot{Institute of Nuclear Physics and University of Mining and Metalurgy,
     Ul. Kawiory 26a, PL-30055 Krakow, Poland
    \label{KRAKOW}}
\titlefoot{Universit\'e de Paris-Sud, Lab. de l'Acc\'el\'erateur
     Lin\'eaire, IN2P3-CNRS, B\^{a}t. 200, FR-91405 Orsay Cedex, France
    \label{LAL}}
\titlefoot{School of Physics and Chemistry, University of Lancaster,
     Lancaster LA1 4YB, UK
    \label{LANCASTER}}
\titlefoot{LIP, IST, FCUL - Av. Elias Garcia, 14-$1^{o}$,
     PT-1000 Lisboa Codex, Portugal
    \label{LIP}}
\titlefoot{Department of Physics, University of Liverpool, P.O.
     Box 147, Liverpool L69 3BX, UK
    \label{LIVERPOOL}}
\titlefoot{LPNHE, IN2P3-CNRS, Univ.~Paris VI et VII, Tour 33 (RdC),
     4 place Jussieu, FR-75252 Paris Cedex 05, France
    \label{LPNHE}}
\titlefoot{Department of Physics, University of Lund,
     S\"olvegatan 14, SE-223 63 Lund, Sweden
    \label{LUND}}
\titlefoot{Universit\'e Claude Bernard de Lyon, IPNL, IN2P3-CNRS,
     FR-69622 Villeurbanne Cedex, France
    \label{LYON}}
\titlefoot{Univ. d'Aix - Marseille II - CPP, IN2P3-CNRS,
     FR-13288 Marseille Cedex 09, France
    \label{MARSEILLE}}
\titlefoot{Dipartimento di Fisica, Universit\`a di Milano and INFN,
     Via Celoria 16, IT-20133 Milan, Italy
    \label{MILANO}}
\titlefoot{Niels Bohr Institute, Blegdamsvej 17,
     DK-2100 Copenhagen {\O}, Denmark
    \label{NBI}}
\titlefoot{NC, Nuclear Centre of MFF, Charles University, Areal MFF,
     V Holesovickach 2, CZ-180 00, Praha 8, Czech Republic
    \label{NC}}
\titlefoot{NIKHEF, Postbus 41882, NL-1009 DB
     Amsterdam, The Netherlands
    \label{NIKHEF}}
\titlefoot{National Technical University, Physics Department,
     Zografou Campus, GR-15773 Athens, Greece
    \label{NTU-ATHENS}}
\titlefoot{Physics Department, University of Oslo, Blindern,
     NO-1000 Oslo 3, Norway
    \label{OSLO}}
\titlefoot{Dpto. Fisica, Univ. Oviedo, Avda. Calvo Sotelo
     s/n, ES-33007 Oviedo, Spain
    \label{OVIEDO}}
\titlefoot{Department of Physics, University of Oxford,
     Keble Road, Oxford OX1 3RH, UK
    \label{OXFORD}}
\titlefoot{Dipartimento di Fisica, Universit\`a di Padova and
     INFN, Via Marzolo 8, IT-35131 Padua, Italy
    \label{PADOVA}}
\titlefoot{Rutherford Appleton Laboratory, Chilton, Didcot
     OX11 OQX, UK
    \label{RAL}}
\titlefoot{Dipartimento di Fisica, Universit\`a di Roma II and
     INFN, Tor Vergata, IT-00173 Rome, Italy
    \label{ROMA2}}
\titlefoot{Dipartimento di Fisica, Universit\`a di Roma III and
     INFN, Via della Vasca Navale 84, IT-00146 Rome, Italy
    \label{ROMA3}}
\titlefoot{DAPNIA/Service de Physique des Particules,
     CEA-Saclay, FR-91191 Gif-sur-Yvette Cedex, France
    \label{SACLAY}}
\titlefoot{Instituto de Fisica de Cantabria (CSIC-UC), Avda.
     los Castros s/n, ES-39006 Santander, Spain
    \label{SANTANDER}}
\titlefoot{Dipartimento di Fisica, Universit\`a degli Studi di Roma
     La Sapienza, Piazzale Aldo Moro 2, IT-00185 Rome, Italy
    \label{SAPIENZA}}
\titlefoot{Inst. for High Energy Physics, Serpukov
     P.O. Box 35, Protvino, (Moscow Region), Russian Federation
    \label{SERPUKHOV}}
\titlefoot{J. Stefan Institute, Jamova 39, SI-1000 Ljubljana, Slovenia
     and Department of Astroparticle Physics, School of\\
     \indent~~Environmental Sciences, Kostanjeviska 16a, Nova Gorica,
     SI-5000 Slovenia, \\
     \indent~~and Department of Physics, University of Ljubljana,
     SI-1000 Ljubljana, Slovenia
    \label{SLOVENIJA}}
\titlefoot{Fysikum, Stockholm University,
     Box 6730, SE-113 85 Stockholm, Sweden
    \label{STOCKHOLM}}
\titlefoot{Dipartimento di Fisica Sperimentale, Universit\`a di
     Torino and INFN, Via P. Giuria 1, IT-10125 Turin, Italy
    \label{TORINO}}
\titlefoot{Dipartimento di Fisica, Universit\`a di Trieste and
     INFN, Via A. Valerio 2, IT-34127 Trieste, Italy \\
     \indent~~and Istituto di Fisica, Universit\`a di Udine,
     IT-33100 Udine, Italy
    \label{TU}}
\titlefoot{Univ. Federal do Rio de Janeiro, C.P. 68528
     Cidade Univ., Ilha do Fund\~ao
     BR-21945-970 Rio de Janeiro, Brazil
    \label{UFRJ}}
\titlefoot{Department of Radiation Sciences, University of
     Uppsala, P.O. Box 535, SE-751 21 Uppsala, Sweden
    \label{UPPSALA}}
\titlefoot{IFIC, Valencia-CSIC, and D.F.A.M.N., U. de Valencia,
     Avda. Dr. Moliner 50, ES-46100 Burjassot (Valencia), Spain
    \label{VALENCIA}}
\titlefoot{Institut f\"ur Hochenergiephysik, \"Osterr. Akad.
     d. Wissensch., Nikolsdorfergasse 18, AT-1050 Vienna, Austria
    \label{VIENNA}}
\titlefoot{Inst. Nuclear Studies and University of Warsaw, Ul.
     Hoza 69, PL-00681 Warsaw, Poland
    \label{WARSZAWA}}
\titlefoot{Fachbereich Physik, University of Wuppertal, Postfach
     100 127, DE-42097 Wuppertal, Germany
    \label{WUPPERTAL}}
\titlefoot{On leave of absence from IHEP Serpukhov
    \label{MILAN-SERPOU}}
\titlefoot{Now at University of Florida
    \label{FLORIDA}}
\endgroup
\clearpage
\headsep 30.0pt
\end{titlepage}
%%%%%%%%%%%%%%%%%%%%%%%%%
%
% Change for the document body
%CD\pagestyle{heading} % for page numbering
\pagenumbering{arabic} % page numbering in number
%CD\renewcommand{\thefootnote}{\fnsymbol{footnote}} % symbolic footnote marks
\setcounter{footnote}{0} %
\large

%   document.tex
%

\def\asmz{$\alpha_s(M_Z)$}
\def\ass{\alpha_s(\sqrt{s})}
\newcommand{\kos}{\ifmmode {{\mathrm K}^{0}_{S}} \else
${\mathrm K}^{0}_{S}$\fi}
\newcommand{\kpm}{\ifmmode {{\mathrm K}^{\pm}} \else
${\mathrm K}^{\pm}$\fi}
\newcommand{\ko}{\ifmmode {{\mathrm K}^{0}} \else
${\mathrm K}^{0}$\fi}
\def\as{$\alpha_s$}
\def\asb{$\alpha_s\sp{b}$}
\def\asc{$\alpha_s\sp{c}$}
\def\asuds{$\alpha_s\sp{udsc}$}
\def\Lam{$\Lambda$}
\def\ZP{Z.\ Phys.\ {\bf C}}
\def\PL{Phys.\ Lett.\ {\bf B}}
\def\PR{Phys.\ Rev.\ {\bf D}}
\def\PRL{Phys.\ Rev.\ Lett.\ }
\def\NP{Nucl.\ Phys.\ {\bf B}}
\def\CPC{Comp.\ Phys.\ Comm.\ }
\def\NIM{Nucl.\ Instr.\ Meth.\ }
\def\Coll{Coll.,\ }
\def\Rmu{$R_3(\mu)/R_3(had)$\ }
\def\Re{$R_3(e)/R_3(had)$\ }
\def\Rmue{$R_3(\mu +e)/R_3(had)$\ }
\def\ee{$e\sp{+}e\sp{-}$}
\def\pslash{\not{\hbox{\kern-2.3pt$p$}}}
\def\thslash{\not{\hbox{\kern-2.3pt$\theta$}}}
\def\eslash{\not{\hbox{\kern-2.3pt$E$}}}

\section {Introduction}

In $e^+e^-$ colliders such as LEP searches for  
new physics can be made with high sensitivity in places where the expected Standard Model
(SM) contributions are small. Events  where all or most particles
are grouped in one direction in space, in a mono-jet type
topology, with one isolated lepton (charged or neutral),
 are a good example of such 
processes. SM extensions related to leptoquark models or single top production via
Flavour Changing Neutral Currents can have such a signature.
 In this paper we report on a 
topological search for events in these two channels.

 Leptoquarks are coloured spin 0 or spin 1 particles with both baryon and 
lepton quantum numbers. These particles are predicted by a variety of extensions
 of the SM, including Grand Unified Theories \cite{Lang}, Technicolor \cite{Dimo} and
composite models \cite{Schremp}. They have electric charges of $\pm$5/3, $\pm$4/3, 
$\pm$2/3 and $\pm$1/3, and decay into a charged or neutral lepton and a quark,
 $L_q \rightarrow l^\pm q$ or $L_q \rightarrow \nu q$. 
Two hypotheses are considered in this paper, one where only
the charged decay mode is possible (charged branching ratio $B = 1.0$),
and one, for leptoquark charges below  $4/3$,
 where both charged and neutral decay modes are equally probable. 
If the leptoquark does not couple to the charged decay mode 
($B = 0$) then these leptoquarks can not be produced singly in $e^+e^-$ 
collisions. 
Leptoquarks may be produced singly or in pairs at $e^+e^-$ colliders. For single 
production, leptoquark mass limits can be set up to almost the kinematical limit. For
this reason only single leptoquark production is considered in this analysis.
 The largest contribution to the production cross-section at LEP is predicted  
to come from processes involving hadrons coming from resolved photons \cite{Donc1}, 
radiated from the incoming beams, which are treated using the Weizacker-Williams 
approximation. The corresponding Feynman diagram is shown 
in figure \ref{fig:lq_diag_new_d2}~a.
Decays of singly produced high mass leptoquarks to a charged lepton
are characterised by a high transverse momentum jet recoiling against a lepton. 
In the decay to a neutrino only the jet is detected.
The initial electron which scatters off the quasi 
real photon is assumed to escape detection down the beam pipe.
 Below the TeV mass range and for couplings of the order of 
the electromagnetic coupling, the leptoquarks should not couple to 
diquarks in order to prevent proton decay. They should also couple chirally to 
either left or right handed quarks but not to both, and mainly diagonally. 
This implies that they should couple to a single leptonic generation and 
to a single quark generation and hence this measurement searches only for 
decays to $e$ and $\nu$.

 The properties of leptoquarks are indirectly constrained by experiments at lower 
energy \cite{Miriam}, by precision measurements of the $Z$ width \cite{eboli}, 
and by direct searches at higher energies \cite{CDF,HERA,H1,LEP}.
The mass of scalar leptoquarks decaying to electron plus jet was constrained
to be above $225~{\rm GeV/c^2}$ using Tevatron data \cite{CDF}. Limits on 
leptoquark masses and couplings were set at HERA using the $e^-p$ 
data \cite{HERA}, giving $M_{Lq}>216-275~{\rm GeV/c^2}$.
An excess of events was found in the $e^+p$ data. The H1 collaboration measured 
a jet-lepton invariant mass of these events ranging
from $187.5~{\rm GeV/c^2}$ up to $212.5~{\rm GeV/c^2}$. Rare processes, 
which are forbidden in the SM, also provide strong bounds
on the $\lambda / m_{Lq}$ ratio \cite{Sacha}, where $\lambda$ is the 
leptoquark-fermion Yukawa type coupling and $m_{Lq}$ is the leptoquark mass.

 In the SM, Flavour Changing Neutral Currents (FCNC) are 
absent at tree level. Neutral currents such 
as  $e^+ e^- \rightarrow t  \bar c (t \bar u)$ can be present at the one loop level, but 
the rates are severely supressed \cite{ganapathi}. 

 Flavour changing vertices are present in many extensions of the SM like 
supersymmetry \cite{Petronzio}, multi-Higgs doublet models \cite{atwood} and
anomalous t-quark production \cite{obraztsov}, which could enhance the
production of top quarks. For instance, in the SM the $t\rightarrow cZ$
 branching ratio is around $10^{-13}$ while in 
the context of a two Higgs doublet model  without natural flavour 
conservation the rates can be higher by more than six orders of
magnitude \cite{atwood}, depending on the chosen parameters. 
 At tree level, single top production is possible via FCNC anomalous couplings 
($e^+ e^- \rightarrow t  \bar c$)  \cite{obraztsov}. The corresponding Feynman diagram is shown 
in figure \ref{fig:lq_diag_new_d2} (b). The $t\rightarrow cZ$ and $t\rightarrow c\gamma$
vertices are described by two anomalous coupling constants $k_{Z}$ and $k_{\gamma}$ respectively.
 Present constraints from LEP--2 data were set \cite{obraztsov} at ($m_t=175~{\rm GeV/c^2}$):

\begin{center}
 $k_{\gamma}^2<0.176$
\\
 $k_{Z}^2<0.533$
\end{center}

 In single top production at LEP, the $t\bar{c} (t\bar{u}$), pair should be produced almost at
rest as the top mass is close to the centre-of-mass energy.
 The top quark decays subsequently to a $b$ quark and a $W$. Only leptonic decays of the $W$
are searched for in this letter. It is an almost background free signature
characterised by one energetic mono-jet and one isolated charged lepton.

\section{The DELPHI Detector and Data Samples}

 A detailed description of the DELPHI detector, its performance, 
the triggering conditions and the readout chain can be found in 
reference \cite{DELPHI}. This analysis relies on the charged particle
detection provided by the tracking system and energy reconstruction 
provided by the electromagnetic and hadronic calorimeters.

 The main tracking detector of DELPHI is the Time Projection Chamber,
which covers the angular range $20^\circ < \theta < 160^\circ$,
where $\theta$ is the polar angle defined with respect to the
beam direction. Other detectors contributing to the track reconstruction
are the Vertex Detector (VD), the Inner and Outer Detectors and the
Forward Chambers. The VD consists of three cylindrical layers
of silicon strip detectors, each layer covering the full azimuthal angle.

 Electromagnetic shower reconstruction is performed in DELPHI using
the barrel and the forward electromagnetic calorimeters, including the
STIC (Small angle TIle Calorimeter), the DELPHI luminosity monitor.

 The energy resolutions of the barrel and forward electromagnetic
calorimeters are parameterized respectively as $\sigma(E)/E = 0.043 \oplus
0.32/\sqrt{E}$ and $\sigma(E)/E = 0.03 \oplus 0.12/\sqrt{E} \oplus
0.11/E$, where $E$ is expressed in {\rm GeV} and the symbol `$\oplus$' implies
addition in quadrature.

 The hadron calorimeter covers both the barrel and forward regions.
It has an energy resolution of $\sigma(E)/E = 0.21 \oplus 1.12/\sqrt{E}$
in the barrel. 

 The effects of experimental resolution, both on the signals and on
backgrounds, were studied by generating Monte Carlo events for the 
possible signals and for the SM processes, and passing them through the 
full DELPHI simulation and reconstruction chain.

 The leptoquark signal was generated for different mass values using the 
PYTHIA generator \cite{Pythia}. The leptoquark production cross-section 
was taken from \cite{Donc}. 

 The $t \bar c (\bar u)$ signal was implemented  in the PYTHIA generator \cite{Pythia} by
producing
a top and c (u) quark pair and allowing the top quark to decay into a b quark and  a
$W$ boson. A singlet colour string was formed between the b and c(u) quarks.

 Bhabha events were simulated with the Berends, Hollik and Kleiss generator \cite{bhabha}.
 PYTHIA was used to simulate 
$e^+ e^- \rightarrow  \tau^+ \tau^- $,
$e^+ e^- \rightarrow Z \gamma$,
$e^+ e^- \rightarrow W^+W^-$,
$e^+ e^- \rightarrow  W^{\pm} e^{\mp} \nu$,
$e^+ e^- \rightarrow Z Z $,
and $e^+ e^- \rightarrow Z e^+ e^-$ events.
In all four fermion channels, studies with the EXCALIBUR generator \cite{excal} 
were also performed.
The two-photon (``$\gamma\gamma$'') physics events were simulated using
the TWOGAM \cite{twogam} generator for quark channels and the Berends, Daverveldt and
Kleiss generator \cite{bdk} for the electron, muon and tau channels.

 Data corresponding to an integrated luminosity of 47.7~pb$^{-1}$ were collected at 
a centre-of-mass energy $\sqrt{s}$ of $183~{\rm GeV}$.

\section{Event Selection}

\begin{figure}[p]
\begin{center}
\vspace{-1cm}
\mbox{\epsfig{file=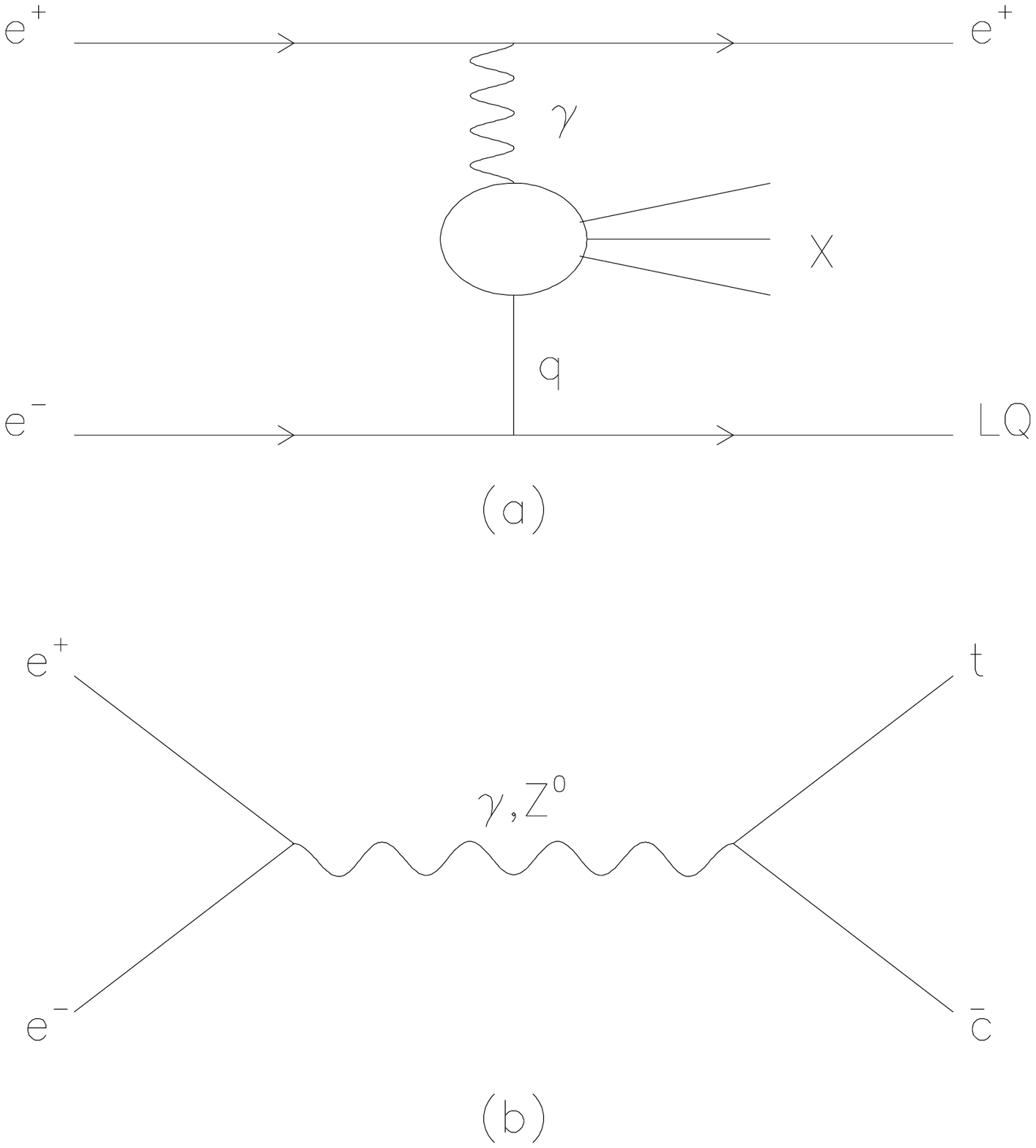,height=0.5\textheight}}
\vspace{-2mm}
\caption{(a) The resolved photon contribution for single leptoquark
production and (b) single top production via FCNC in $e^+e^-$ collisions.}
\label{fig:lq_diag_new_d2}
\end{center}
%KEYS{searches;leptoquarks;val0}
\end{figure}

 This analysis looks for events with one energetic mono-jet. Leptoquark decays 
to a charged lepton and $t\bar{c}$ decays also require an isolated charged lepton. 
The recoil electron in figure \ref{fig:lq_diag_new_d2} (a)
is expected to pass undetected down the beam pipe while
the products of the recoil (X) in figure \ref{fig:lq_diag_new_d2} (a) and the c-quark
 in figure \ref{fig:lq_diag_new_d2} (b) are of low energy
and are absorbed into the mono-jet or lepton.

 Charged particles were considered only if they had
momentum greater than 0.1 {\rm GeV/c} and
impact parameters in the transverse plane and in the beam
direction below 4 cm and 10 cm respectively.
Neutral clusters were defined as energy depositions in the
calorimeters unassociated with charged particle tracks.
All electromagnetic (hadronic) neutrals of energy above 100~{\rm MeV} 
(1~{\rm GeV}) were selected. In 
the present analysis the minimum required charged multiplicity was six.

 Charged particles were considered isolated if, in 
a double cone centred on their track with internal and external half 
angles of $5^\circ$ and $25^\circ$, the total energy associated to charged 
and neutral particles was below 1 {\rm GeV} and 2 {\rm GeV} respectively.
The energy of the particle was redefined as the sum of the energies of
all the charged and neutral particles inside the inner cone. This energy
was required to be greater than 4 {\rm GeV}. No other charged particle was allowed
inside the inner cone.

 Energy clusters in the electromagnetic calorimeters were
considered to be from photons if there were no tracks pointing to the cluster,
there were
no hits inside a 2$^\circ$ cone in more than one layer of the
Vertex Detector and if at least
90$\%$ of any hadronic energy was deposited in the
first layer of the hadron calorimeter. Photons were considered to be isolated if,
 in a double cone centred on the cluster and having internal and external half angles of
$5^\circ$ and $15^\circ$, the total energy deposited was less than 1 {\rm GeV}. 
The energy of the photon was redefined as the sum of the energies
of all the particles inside the inner cone and
no charged particles above 250~${\rm MeV/c}$ were allowed inside this cone.

 All charged and neutral particles (excluding any isolated charged lepton, if present)
were forced into one jet using the Durham jet algorithm \cite{Durham}. The jet was 
classified as charged if it contained at least one charged particle.

 A detailed description of the basic selection criteria can be found in 
reference \cite{ourpaper}. Isolated charged particles were identified as 
electrons if there were no associated hits in the muon chambers, 
if the ratio of the energy measured in the electromagnetic
calorimeters, E, to the momentum measured in the tracking chambers, p, was
larger than 0.2 and if the energy deposited in the electromagnetic
calorimeters by the lepton candidate was at least 90\% of the total energy
deposited in both electromagnetic and hadronic calorimeters.

The following criteria were applied to the events (level 1):

\begin{itemize}

\item
 the total visible energy was required to be larger than 0.2$\sqrt{s}$;

\item
  events with isolated photons were rejected;

\item 
 the momentum of the monojet was required to be larger than 10~${\rm GeV/c}$;

\item 
 in channels with one isolated charged particle its momentum had to be
greater than 10~${\rm GeV/c}$; for the leptoquark search exactly one 
isolated charged particle was required in 
the event; for the FCNC search at least one charged isolated particle
 was required.

\end{itemize}

After this selection, more specific criteria were applied (level 2):

\begin{itemize}

\item
 Events were required to have only one jet with the Durham resolution variable 
($y_{cut}$) \cite{Durham} in the transition from one to two jets smaller than 0.09.
 
\item
 The monojet polar angle had to be between $30^\circ$ ($20^\circ$) and 
$150^\circ$ ($160^\circ$) for the leptoquark search (for the FCNC search).
 
\item
 The ratio between the monojet electromagnetic energy and its total 
energy had to be smaller than 0.9. This removes most Bhabha events.

\item 
 The sum of the transverse momentum of the charged particles in the jet
 (relative to the event thrust axis) normalized to the total visible momentum 
had to be lower then 0.17. This cut reduces the contamination from semileptonic
 decays of $WW$ pairs.

\end{itemize}
 
\begin{figure}[p]
\begin{center}
\vspace{-1cm}
\mbox{\epsfig{file=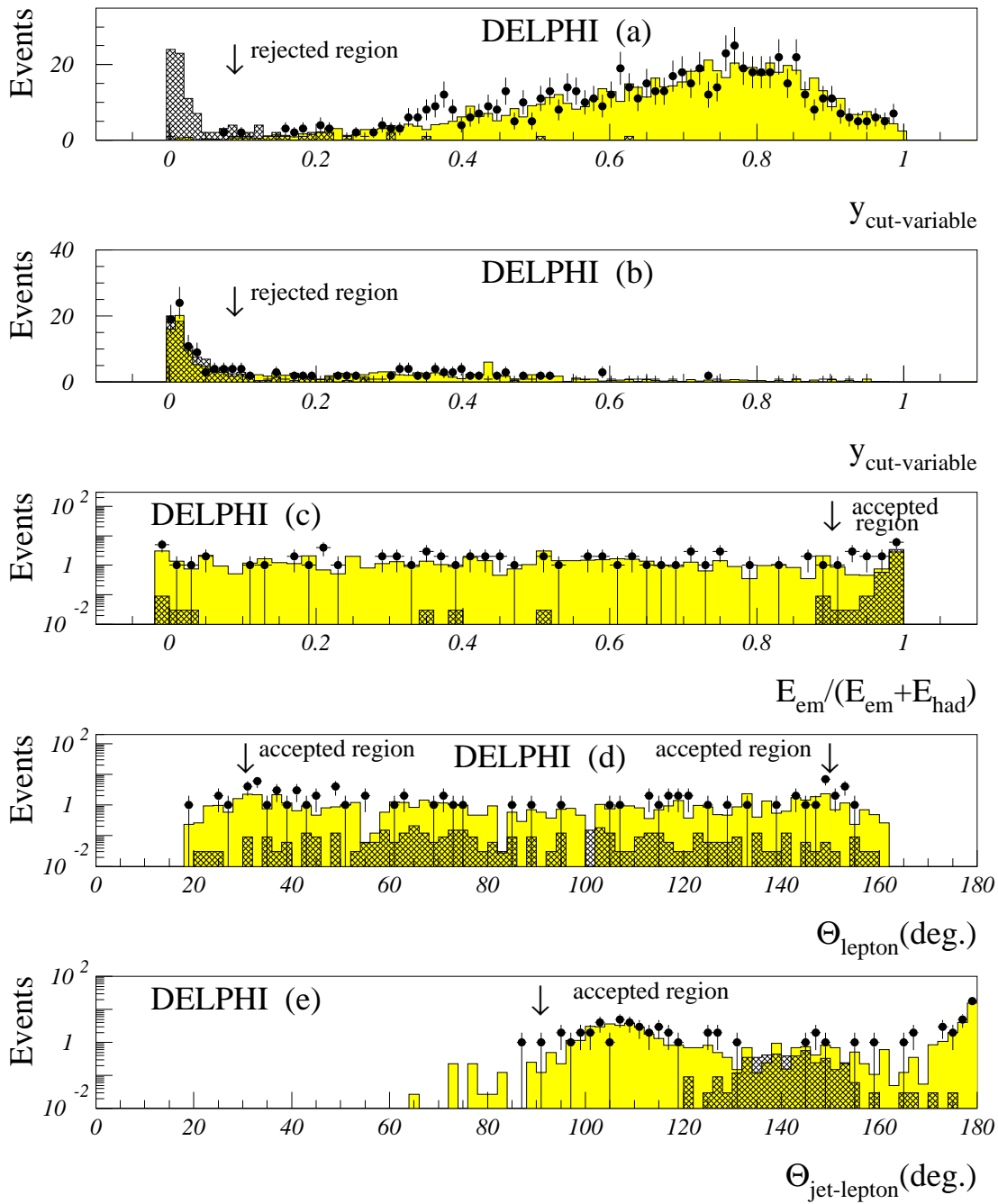,height=0.8\textheight}}
\vspace{-2mm}
\caption{Leptoquark search: (a) the $y_{cut}$ variable distribution for neutral decays (Level 1),
 (b) the $y_{cut}$ variable distribution for charged decays (Level 1), (c) the ratio between the 
energy deposited in the electromagnetic calorimeters by the lepton candidate and the total energy deposited in both
electromagnetic and hadronic calorimeters (Level 2), (d) the lepton polar 
angle (Level 2) and (e) the angle between the jet and
the lepton (Level 2). The dots show the data and the shaded region shows the SM simulation.
The dark region is the expected signal behaviour for a leptoquark mass of 120~${\rm GeV/c^2}$.
The vertical arrows show the cut used to select events. The accepted or rejected region is also shown.}
\label{fig:new_fig2_lepto_d2}
\end{center}
%KEYS{searches;leptoquarks;val0}
\end{figure}

In the case of the leptoquark neutral decays the  
$y_{cut}$ criterion is the most effective for distinguishing
signal from background.    
This is illustrated in Figure \ref{fig:new_fig2_lepto_d2} (a) 
where the dots show the data, the shaded region the SM simulation
and the dark region the expected signal behaviour.  The same
distributions are shown in Figure 
\ref{fig:new_fig2_lepto_d2} (b) for the leptoquark charged decays.

Additional criteria (level 3) were applied in order to reduce the contamination from
background events, mostly $q \bar q$ and $WW$. These criteria were different for
the different channels:

\begin{description}[1em]

\item[-]
 For the leptoquark charged decay mode it was required that:

\begin{description}[1em]

 \item[(i)] the lepton was identified as an electron and its polar angle had to be 
between $30^\circ$ and $150^\circ$;
 
\item[(ii)] the angle between the electron and the monojet had to be 
larger than $90^\circ$.

\end{description}

\item[-]
 For the leptoquark neutral decay mode,
where the contamination of $q\bar{q}$ is higher, all particles were 
also forced into two
jets, and the following additional criteria were applied: 

\begin{description}[1em]
 
\item[(i)] the angle between the two jets had to be smaller than $155^\circ$;
 
\item[(ii)] the momentum of the second jet had to be smaller than 10~${\rm GeV/c}$, whenever 
the angle between the two jets was larger than $60^\circ$.

\end{description}

\item[-] For the single top production:

\begin{description}[1em]
 
\item[(i)] the polar angle of the most energetic lepton had to be between $20^\circ$  and
$160^\circ$, and the angle between the lepton and the monojet had to be between 
$15^\circ$ and $165^\circ$;
 
\item[(ii)] events with a B hadron decay were selected by requiring the b-tag
 variable \cite{vanina} to be below 0.06;

\item[(iii)] the polar angle of the missing momentum had to be between $20^\circ$ and
$160^\circ$.

\end{description}

\end{description}

 In table \ref{tab:statistics} the number of events which survived 
the different levels of selection is shown, together with the expected SM
background. The $WW$ and $q \bar q$ events are the main source of 
background. At level 3 the expected background contribution from $WW$ and 
$q \bar q$ events is: for the leptoquark neutral decay mode,
 $0.12 \pm 0.12$ and $0.46 \pm 0.33$ respectively; for the leptoquark 
charged decay mode $0.12 \pm 0.12$ and $0.69 \pm 0.4$ respectively; for 
the FCNC $0.49 \pm 0.25$ and $0.23 \pm 0.23$ respectively.
 Figure \ref{fig:new_fig2_lepto_d2} (c) shows (at level 2), for the leptoquark search, 
the ratio between the energy deposited in the electromagnetic
calorimeters by the lepton candidate and the total energy
deposited in both electromagnetic and hadronic calorimeters, (d) the lepton polar angle and (e) 
the angle between the jet and the lepton.
 The dots show the data and the shaded region shows the SM simulation. 
The dark region is the expected signal behaviour for a 120${\rm GeV/c^2}$ leptoquark mass.
No upper bound was imposed in the jet lepton angle to allow good signal 
efficiency up to threshold (where the jet and the lepton are essentially 
back to back). However the selection on figure \ref{fig:new_fig2_lepto_d2} (c) 
removes almost all the SM background on figure \ref{fig:new_fig2_lepto_d2} (e).

\begin{figure}[p]
\begin{center}
\vspace{-1cm}
\mbox{\epsfig{file=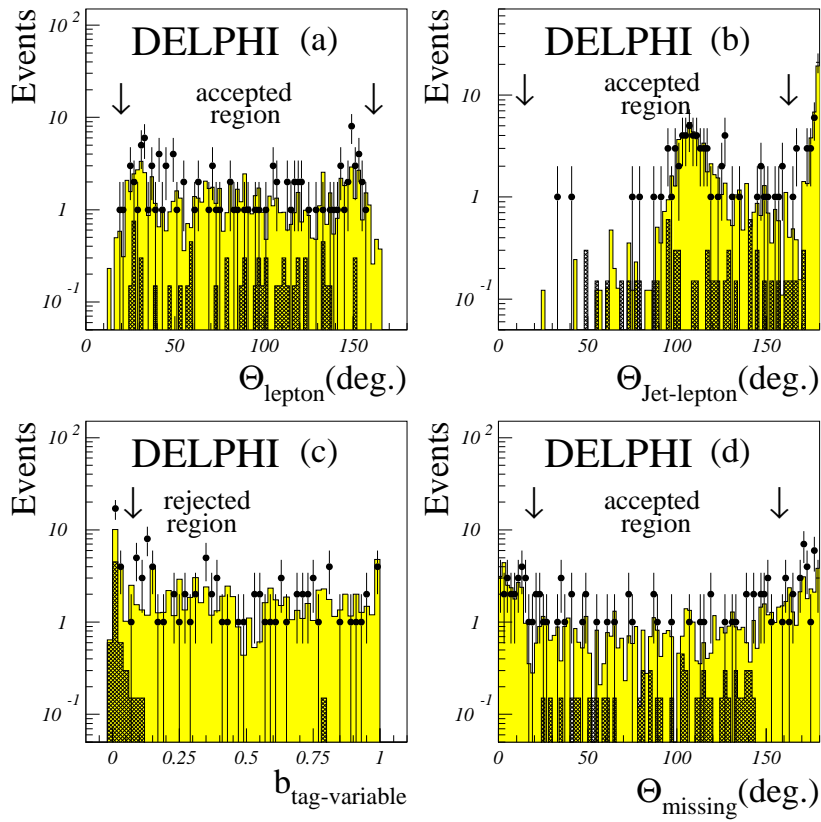,height=0.5\textheight}}
\vspace{-2mm}
\caption{FCNC search: (a) the lepton polar angle, (b) the jet-lepton angle, (c) the b-tag
variable (see text) and (d) the missing momentum polar angle. The dots show the data and the shaded
region shows the SM simulation. The dark region is the expected signal behaviour.
The vertical arrows show the cut used to select events. The accepted or rejected region 
is also shown.}
\label{fig:new_fig3_fcnc_d2}
\end{center}
%KEYS{searches;leptoquarks;val0}
\end{figure}

 Figure \ref{fig:new_fig3_fcnc_d2} shows (at level 2), for the FCNC search, (a) the lepton polar
angle, (b) the jet-lepton angle, (c) the b-tag variable \cite{vanina} and (d) the missing
momentum polar angle. The dots show the data and the shaded region shows the SM simulation. 
The dark region is the expected signal behaviour.
A good agreement is observed.\\

\begin{table}[hbt]
\begin{center}
%\vskip 0.45 cm
\begin{tabular}{|l||c|c||c|}
\hline
       & \multicolumn{2}{c||} {\bf Leptoquark} 
& \multicolumn{1}{c|} {\bf FCNC} \\
\cline{2-4}
        & {$Charged Decay$} & {$Neutral Decay$} & {$Charged Decay$} \\
        & {$Data~~(SM)$} & {$Data~~(SM)$} & {$Data~~(SM)$} \\
\hline
\hline
 {\bf Level 1} &~537~(501$\pm12$)~&~3159~(2917$\pm28$)~&~572~(542$\pm12$)~\\ 
 {\bf Level 2} &~~76~(~64$\pm~4$)~&~~~~4~(~2.6$\pm.7$)~&~101~(~96$\pm~5$)~\\ 
 {\bf Level 3} &~~~1~(1.1$\pm.5$)~&~~~~1~(~1.0$\pm.4$)~&~~~0~(1.1$\pm.4$)~\\ 
\hline
\end{tabular}
\end{center}
\caption[]{ Number of selected data events and expected SM contributions for the
charged and neutral decay modes at different levels of selection criteria.}
\label{tab:statistics}
\end{table}

\section{Results for Leptoquarks}

 Only first-generation leptoquarks were searched for in this analysis 
($L_q \rightarrow e^\pm q$, $L_q \rightarrow \nu_{e} q$).
 As discussed previously, the highest contribution to the production 
cross-section relevant for this search comes from the resolved photon 
contribution. The Gl\"uck-Reya-Vogt parameterization \cite{GRV} of the parton 
distribution was used. Since the photon has different u-quark and d-quark 
contents and the production cross-section is proportional to $(1+q)^2$
(where $q$ is the leptoquark charge), leptoquarks of charge $q=-1/3$ and
$q=-5/3$ (as well as leptoquarks of charge  $q=-2/3$ and  $q=-4/3$)
 have similar production cross-sections \cite{Donc}.  The cross-sections
used here were calculated within the assumption above.

\subsection{Charged Decay Mode}

In this channel one event was found in the data at 
$\sqrt{s} = 183~{\rm GeV}$ and the expected SM background was $1.1\pm0.5$.

 The leptoquark invariant mass estimated from the energies and directions of the
 jet and lepton is 89.9~${\rm GeV/c^2}$.
 The mass resolution ranges from 15~${\rm GeV/c^2}$ to 25~${\rm GeV/c^2}$
for leptoquark masses from 100~${\rm GeV/c^2}$ up to the kinematical
limit.

 Within the low statistics there is good agreement between data and SM predictions. 

The efficiency was found to be between 22\% and 30\% 
for leptoquark masses in the range from 100~${\rm GeV/c^2}$ up to the kinematic
limit.

\subsection{Neutral Decay Mode}

In this channel one event was found and the expected SM 
background was $1.0\pm0.4$.

 The leptoquark invariant mass estimated from the monojet transverse momentum 
 is 72.1~${\rm GeV/c^2}$.
 The mass resolution ranges from 20~${\rm GeV/c^2}$ to 34~${\rm GeV/c^2}$
for leptoquark masses from 100~${\rm GeV/c^2}$ up to the kinematical
limit.
 Within the low statistics there is good agreement between data and SM predictions. 

The efficiency was found to be between 20\% and 41\% 
for leptoquark masses in the range from 100~${\rm GeV/c^2}$ up to the kinematic
limit.

\subsection{Leptoquark Mass and Coupling Limits}

\begin{figure}[p]
\begin{center}
\vspace{-1cm}
\mbox{\epsfig{file=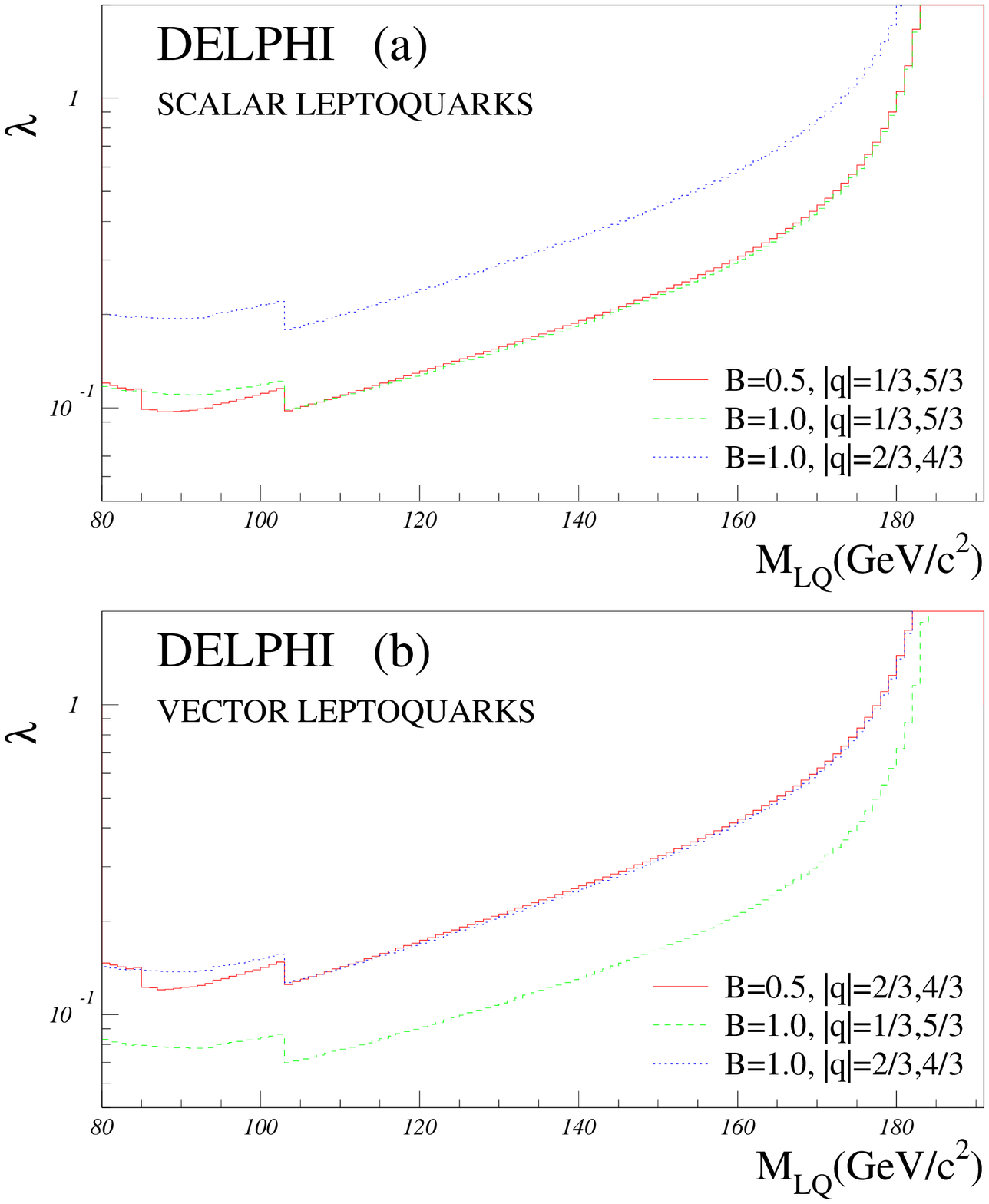,height=0.7\textheight}}
\vspace{-2mm}
\caption{95\% confidence level upper limits on the coupling $\lambda$ 
as a function of the leptoquark mass for (a) scalar and (b) vector 
leptoquarks (B is the branching ratio of the leptoquark to charged leptons and q is the
leptoquark charge). }
\label{fig:lq_lims_new_d2}
\end{center}
%KEYS{searches;leptoquarks;val0}
\end{figure}

 Limits were set on the leptoquark coupling parameter $\lambda$ \cite{Donc1}.
These limits, which depend on the leptoquark mass, are shown
 in figure~\ref{fig:lq_lims_new_d2} for both scalar and vector leptoquarks
of different types and for charged decay branching ratios 
$B = 1$ and $B = 0.5$. For $B = 1$ the invariant mass plot for the charged
decay mode was used to set the limits. For $B = 0.5$ the invariant mass 
plots of the charged and the neutral decay modes were combined to set the limits. 
Different values of the charged decay branching ratio B, although theoretically not
motivated, would imply similar limits.

The lower limits at 95\% confidence level on the mass of a first 
generation leptoquark for a coupling parameter
 $\lambda=\sqrt{4 \pi \alpha_{em}}$ are given in table \ref{tab:limits}, where
different leptoquark types and branching ratios are considered \cite{rizzo}.
 These limits are expected to change at the level of some percent 
depending on the different theoretical predictions for the total 
production cross section \cite{Papa}.

\begin{table}[hbt]
\begin{center}
%\vskip 0.45 cm
\begin{tabular}{|l||r|r||r|r|}
\hline
       & \multicolumn{2}{c||} {$B = 0.5$}
& \multicolumn{2}{c|} {$B = 1.0$} \\
\cline{2-5}
       & {$|q|$=1/3} & {$|q|=2/3$} & {$|q|=1/3,5/3$} & {$|q|=2/3,4/3$} \\
\hline
\hline
 scalar & 161  & - & 161 & 134   \\ 
 vector & -  & 149 & 171  & 150   \\ \hline
\end{tabular}
\end{center}
\caption[]{ Lower limits (in ${\rm GeV/c^2}$) at 95\% confidence level on the
the mass of a first generation leptoquark for a coupling parameter of
$\lambda$={$\sqrt{4 \pi \alpha_{em}}$}.}  
\label{tab:limits}
\end{table}

\section{Results for Top-Charm FCNC}

In the present analysis no events were found
while the expected SM background is $1.1\pm0.4$. The detection efficiency, 
including the $W$ leptonic branching ratio, is $(11.5 \pm 2.0 )\%$.

With the present luminosity of 47.7 pb$^{-1}$, an upper limit on the 
$e^+ e^- \rightarrow t \bar c$  Flavour Changing Neutral Current
 total cross-section can be set at 0.55~pb (95\% confidence level).  

 This value can be translated into a limit on the anomalous coupling 
constants $k_{\gamma}$ and $k_{Z}$, according to the parametrization 
described in reference \cite{obraztsov}. It was assumed that both 
channels $e^+ e^- \rightarrow t  \bar c$  and $e^+ e^- \rightarrow t  \bar u$
 contributed to the total cross-section.  With a luminosity
of 47.7 pb$^{-1}$ the 95\% confidence level upper limit on
$k_{\gamma}$ is 2, for a $k_{Z}$ value of zero, and the
corresponding upper limit on $k_{Z}$ is 1.5, for a $k_{\gamma}$
value of zero.  The results are not yet
competitive with other experimental results 
\cite{CDF_tc}.

\section{Conclusions}

A search for first generation leptoquarks was performed 
using the data collected by the DELPHI detector at $\sqrt{s}=183~{\rm GeV}$. 
Both neutral and charged decay modes of scalar and vector leptoquarks were 
searched for. No evidence for a signal was found in the data. Limits on 
leptoquark masses were set ranging from $134~{\rm GeV/c^2}$ 
to $171~{\rm GeV/c^2}$ at 95\% confidence level, 
assuming electromagnetic type couplings.

 A search for $t \bar{c}$ flavour changing neutral currents was also performed. No signal was found in the data. A limit on the FCNC cross-section 
was set at $0.55$~pb (95\% confidence level).

%%%\newpage
\subsection*{Acknowledgements}
\vskip 3 mm
%=========================================================================%

We would like to thank M. Doncheski and C. Papadopoulos for the very useful
discussions on the leptoquark production. We would also like to thank D. Atwood,
 L. Reina and A. Soni for the on-going discussion relative to the two Higgs 
doublet model. \\
%         Created on 12-FEB-1998 by dimartino
%-------------------------------------------------------------------
%\subsection*{Acknowledgements}
%\vskip 3 mm
 We are greatly indebted to our technical 
collaborators, to the members of the CERN-SL Division for the excellent 
performance of the LEP collider, and to the funding agencies for their
support in building and operating the DELPHI detector.\\
We acknowledge in particular the support of \\
Austrian Federal Ministry of Science and Traffics, GZ 616.364/2-III/2a/98, \\
FNRS--FWO, Belgium,  \\
FINEP, CNPq, CAPES, FUJB and FAPERJ, Brazil, \\
Czech Ministry of Industry and Trade, GA CR 202/96/0450 and GA AVCR A1010521,\\
Danish Natural Research Council, \\
Commission of the European Communities (DG XII), \\
Direction des Sciences de la Mati$\grave{\mbox{\rm e}}$re, CEA, France, \\
Bundesministerium f$\ddot{\mbox{\rm u}}$r Bildung, Wissenschaft, Forschung 
und Technologie, Germany,\\
General Secretariat for Research and Technology, Greece, \\
National Science Foundation (NWO) and Foundation for Research on Matter (FOM),
The Netherlands, \\
Norwegian Research Council,  \\
State Committee for Scientific Research, Poland, 2P03B06015, 2P03B03311 and
SPUB/P03/178/98, \\
JNICT--Junta Nacional de Investiga\c{c}\~{a}o Cient\'{\i}fica 
e Tecnol$\acute{\mbox{\rm o}}$gica, Portugal, \\
Vedecka grantova agentura MS SR, Slovakia, Nr. 95/5195/134, \\
Ministry of Science and Technology of the Republic of Slovenia, \\
CICYT, Spain, AEN96--1661 and AEN96-1681,  \\
The Swedish Natural Science Research Council,      \\
Particle Physics and Astronomy Research Council, UK, \\
Department of Energy, USA, DE--FG02--94ER40817. \\
%=========================================================================%

\newpage
           
%%%%% {References}

%\begin{figure}[p]
%\begin{center}
%\vspace{-1cm}
%\mbox{\epsfig{file=lq_kzkg_verynew_d2.eps,height=0.7\textheight}}
%\vspace{-2mm}
%\caption{95\% confidence level upper limit on the $k_{\gamma}$-$k_{Z}$ plane.
% The shadowed region corresponds to the CDF exclusion 
%region (obtained from reference \cite{CDF_tc}). The hatched region
% (overlaped with the shadowed one) corresponds
%to the DELPHI exclusion region obtained with the luminosity of 47.7 pb$^{-1}$.
%}
%\label{fig:lq_kzkg_verynew_d2}
%\end{center}
%%KEYS{searches;leptoquarks;val0}
%\end{figure}   


\begin{thebibliography}{99}                                                     
\bibliographystyle{unsrt}                                     



\bibitem{Lang} P. Langacker, Phys. Rep. {\bf 72} (1981) 185.

\bibitem{Dimo} See for example S. Dimopoulos, Nucl. Phys. {\bf B168} (1981) 69.

\bibitem{Schremp} See for example B. Schremp and F. Schremp, Phys. Lett. 
{\bf B153} (1985) 101.

\bibitem{Donc1} M. Doncheski and S. Godfrey, Phys. Rev. {\bf D49}
(1994) 6220.

\bibitem{Miriam} O. Shanker, Nucl. Phys. {\bf B204} (1982) 375;
W. Buchmuller and D.Wyler, Phys. Lett. {\bf B177} (1986) 377;
J.L. Hewett and T.G. Rizzo, Phys. Rev. {\bf D36} (1987) 3367;
M. Leurer, Phys. Rev. {\bf D49} (1994) 333 and
Phys. Rev. {\bf D50} (1994) 536.


\bibitem{eboli}
J.K. Mizukoshi, O.J.P. Eboli and M.C. Gonzalez-Garcia, CERN-TH 7508/94 
(1994); G. Bhattacharya, J. Ellis and K. Sridhar, CERN-TH 7280/94 (1994). 


\bibitem{CDF}
CDF Coll., F. Abe et al., hep-ex/9708017; D0 Coll., B. Abbott et
al., Fermilab-Pub-97/252-E (hep-ex/9707033).


\bibitem{HERA}
ZEUS Coll., M. Derrick et al., Phys. Lett. {\bf B306} 1993 (173);
H1 Coll., I. Abt et al., Nucl. Phys. {\bf B396} (1993) 3.


\bibitem{H1}
H1 Coll., C. Adloff et al., DESY 97-024 (hep-ex/9702012); ZEUS Coll.,
J. Breitweg et al. DESY 97-025 (hep-ph/9702015); updated analysis see 
B. Straub, talk presented at LP'97 Symposium, Hamburg, July 1997.


\bibitem{LEP}
ALEPH Coll., D. Decamp et al., CERN PPE/91-149.
DELPHI Coll., P. Abreu et al., Phys. Lett. {\bf B316} (1993) 620;
L3 Coll., B. Adeva et al., Phys. Lett. {\bf B261} (1991) 169;
OPAL Coll., G. Alexander et al., Phys. Lett. {\bf B263} (1991) 123.


\bibitem{Sacha}
 S. Davidson et al., Z. Phys. {\bf C61} (1994) 613.




\bibitem{ganapathi}
V. Ganapathi, T. Weiler, E.Laermann, I. Schmitt, and P.M. Zerwas, Phys. Rev.
{\bf D27}, (1983) 579; A. Axelrod, Nucl. Phys. {\bf B209} (1982) 349;
 G. Eilam, J.L. Hewett, A. Soni, Phys. Rev. {\bf D44} (1991) 1473; 
B. Grzadkowski, J.F. Gunion, and P. Krawczyk, Phys. Lett. {\bf B268} (1990) 106.


\bibitem{Petronzio}
G.M. Divitiis, R. Petronzio and L. Silvestrini, hep-ph/9704244.


\bibitem{atwood}
D. Atwood, L. Reina and A. Soni, hep-ph/9506243, SLAC-PUB-95-6927.


\bibitem{obraztsov}
V.F. Obraztsov, S.R. Slabospitsky and O.P. Yuschchenko, IHEP-97-79 (hep-ph/9712394).



\bibitem{DELPHI} 
DELPHI Coll., P. Aarnio et al., Nucl. Instr. Meth. {\bf A303} (1991) 233; \\
DELPHI Coll., P. Abreu et al., Nucl. Instr. Meth. {\bf A378} (1996) 57.



\bibitem{Pythia}
 T. Sj\"ostrand, Comp. Phys. Comm. {\bf 82} (1994) 74; \\
 T. Sj\"ostrand, Pythia 5.7 and Jetset 7.4, CERN-TH/7112-93.


\bibitem{Donc} M. Doncheski and S. Godfrey, Phys. Lett. {\bf B393}
(1997) 355.


\bibitem{bhabha}
F.A.Berends, W. Hollik and R. Kleiss, Nucl. Phys. {\bf B304} (1998) 712.


\bibitem{excal}
F.A.~Berends, R.~Pittau, R.~Kleiss, Comp. Phys. Comm. {\bf 85} (1995) 437.


\bibitem{twogam}
S. Nova, A. Olchevski and T. Todorov,
``TWOGAM, a Monte Carlo event generator for two photon physics",
DELPHI Note 90-35 PROG 152.


\bibitem{bdk}
F.A.Berends, P.H.Daverveldt, R. Kleiss, Comp. Phys. Comm. {\bf 40} (1986) 271.



\bibitem{Durham}
 S. Catani et al.,  Phys. Lett. {\bf B269} (1991) 432.
% S. Bethke et al., Nucl. Phys. {\bf B370} (1992) 310.

  
\bibitem{ourpaper}
DELPHI Coll., P. Abreu et al., Phys. Lett. {\bf B393} (1997) 245.


\bibitem{vanina} G. Borisov, C. Mariotti, DELPHI 97-16 PHYS672;
DELPHI Coll., P. Abreu et al., Nucl. Inst. Meth. {\bf A378} (1996) 57.

 
\bibitem{GRV} M. Gluck et al., Phys. Rev. {\bf D46} (1992) 1973 and
Phys. Rev. {\bf D45} (1992) 3986.


\bibitem{rizzo}
J.L. Hewett and T.G. Rizzo, hep-ph/9703337.


\bibitem{Papa}
C. Papadopoulos, hep-ph/9703372.


\bibitem{CDF_tc}
CDF Coll., Fermilab-Pub-97/270-E (1997).



\end{thebibliography}
\end{document}